\documentclass[fleqn,twoside]{article}
\usepackage{espcrc2}
\usepackage{epsfig}
\usepackage{psfrag}

\newcommand{\bel}[1]{\begin{equation}\label{#1}}
\newcommand{\be}{\begin{equation}}
\newcommand{\ee}{\end{equation}}
\newcommand{\Tr}{{\rm Tr}}
\newcommand{\sign}{{\rm sign}}

\begin{document}
\title{
\vspace{-3cm}
\rightline{\small KANAZAWA-02-24}
\rightline{\small ITEP-LAT/2002-13}
\vspace{2cm}
Confining string and P-vortices in the indirect $Z(2)$ projection of $SU(2)$ lattice gauge theory
\thanks{Talk presented by A.~V.~Kovalenko at Lattice 2002 symposium, Boston}} \author{V.~G.~Bornyakov\address[KU]{Department of Physics, Kanazawa University, Kanazawa~920-11,
Japan}$^{,}$\address[ITEP]{ITEP, B.~Cheremushkinskaya~25, Moscow~117259, Russia},
A.~V.~Kovalenko\addressmark[ITEP],
M.~I.~Polikarpov\addressmark[ITEP], D.~A.~Sigaev\addressmark[ITEP]}

\begin{abstract}
We study the distribution of P-vortices near the confining string in the indirect $Z(2)$ projection
of $SU(2)$ lattice gauge theory. It occurs that the density of vortices is constant at large distances
and strongly suppressed near the line connecting the test quark-antiquark pair.
This means that the condensate of 
P-vortices is broken inside the confining string. We also find that the 
width of the P-vortex density distribution  is proportional to the logarithm of the 
distance between the quark and antiquark.
\end{abstract}
\maketitle

\section{Introduction}
Two popular confinement mechanisms are the monopole confinement~\cite{DualMeissner,AbProjection}
and the magnetic vortex confinement~\cite{AA}. The numerical study of these mechanisms on the lattice shows that
both of them have attractive features, e.~g. the abelian dominance~\cite{AbDom}, an explanation of the Casimir
scaling~\cite{AB}, etc. 
Studies of the confining string in the maximal abelian projection~\cite{AD}
clearly demonstrate that
QCD vacuum behaves like the dual superconductor. Below we present results of an analogous study of the
confining string in terms of P-vortices in the indirect $Z(2)$ projection of $SU(2)$ lattice gauge theory.

We use the standard definition~\cite{AB} for P-vortices in the indirect $Z(2)$ projection. After
fixing the maximal abelian gauge we maximize the functional $R$ and define the link variables $Z_{x\mu}$:
$$R=\sum_{x,\mu}\cos^2\theta_{x\mu},\;\; Z_{x\mu}=\sign(\cos\theta_{x\mu}),$$
where 
$\cos\theta_{x\mu}=\frac12\Tr\, u_{x\mu}$, $u_{x\mu}$ is the abelian link matrix
obtained from $SU(2)$ link matrix  $U_{x\mu}$ by the abelian projection.

A plaquette is pierced by a P-vortex lying on the dual lattice 
if $Z(2)$ plaquette $P_{x,\mu\nu} \equiv Z_{x\mu} Z_{x+\mu,\nu} Z^\dagger_{x+\nu,\mu} Z^\dagger_{x\nu} = -1$.

\section{Vortices near the confining string}
We consider the quantity that measures the dependence of the vortex density  
on the distances $r_\perp$ from the $q\bar{q}$ axis and $r_\parallel$ along
this axis
\bel{quan}Q(r_\perp,r_\parallel) = \frac{{<}W V(r_\perp,r_\parallel){>}}{{<}W{>}},\ee
where W is a Wilson loop and vortex detector $V(r_\perp,r_\parallel) = 
(1-P_{x,\mu\nu})/2$. 
The similar quantity was investigated for the distribution of monopole currents near the confining string~\cite{AD}.
Our calculations were performed on $SU(2)$ vacuum configurations in the indirect $Z(2)$ gauge,
which was fixed using the simulated annealing algorithm~\cite{AF}. Below we present results obtained on $50$
statistically independent configurations generated on $24^4$ lattice for $\beta=2.5$.

\begin{figure}
\psfrag{RRRRRR}{$r_\parallel$, fm}
\psfrag{rrrrrr_rrrrrr}{$r_\perp$, fm}
\psfrag{Q}{$\frac{Q}{a^2}$}

\epsfig{file=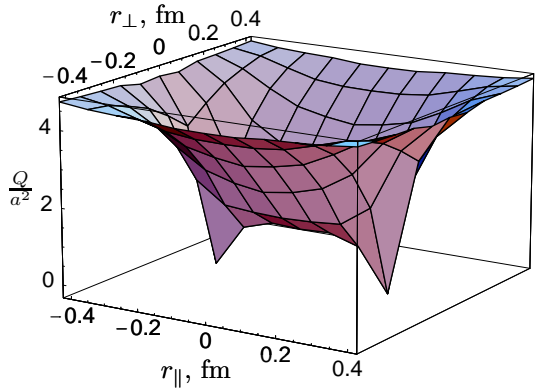,width=7.51cm}
\caption{The dependence of $Q/a^2$ on $r_\perp$ and $r_\parallel$.}
\label{3d1}
\end{figure}

Obviously there is no correlation between $W$ and $V(r_\perp,r_\parallel)$ 
at large distances $r_\perp$, $Q(\infty,r_\parallel)={<}V{>}$,
which is the value of the string condensate~\cite{maxim}.
However, the vortex density decreases near the confining string.
This effect is illustrated in Fig.~\ref{3d1}. The minima on Fig.~\ref{3d1} 
correspond to the positions of the test quark and antiquark.

\begin{figure}[!ht]
\psfrag{rrrrrr}{$r_\perp$, fm}
\psfrag{QQQQQQ}{$Q/a^2\rm,\;\;fm^{-2}$}

\epsfig{file=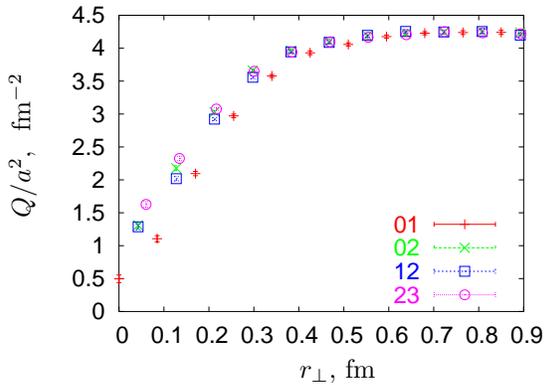,width=7.51cm}
\caption{$Q/a^2$ as the function of $r_\perp$ for the
Wilson loop $11\times5$. 
Notation `01' means that the detecting plaquette lies in
the plane `01'.
}
\label{all_vort}
\end{figure}
\begin{figure}
\psfrag{rrrrrr}{$r_\perp$, fm}
\psfrag{QQQQQQ}{$Q/a^2\rm,\;\;fm^{-2}$}

\epsfig{file=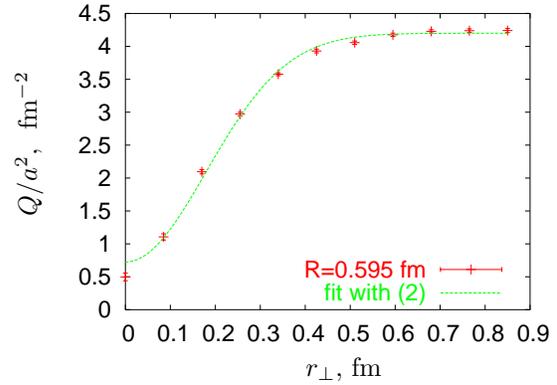,width=7.51cm}
\caption{The fit of the string profile by expression~(\ref{fit}) for the 
Wilson loop $11\times5$. The data are for plaquettes `01'.}
\label{11x5}
\end{figure}
In Fig.~\ref{all_vort} we show the dependence of the vortex density 
$Q(r_\perp,r_\parallel)/a^2$ on the distance $r_\perp$ for $r_\parallel$ 
fixed in the middle of the $q\bar{q}$ axis and the vortex detector $V$ lying in various 
planes.
The Wilson loop has temporal direction '0' and spatial direction '1'.
We see stronger suppression of $Q$ for the plane $(0,1)$ parallel to the Wison loop.
The normalization $1/a^2$ is due to dimensional reasons~\cite{AB}, 
the density of P-vortices should scale as $a^2$, $a$ being the lattice spacing. 

\section{The radius of the confining string}

We employ  $Q(r_\perp,r_\parallel)$ to evaluate  the radius of the string in terms of P-vortices.

The Gaussian form fitting function is used to fit our data:
\bel{fit} Q(r_\perp) =  A \exp\left[-\left(\frac {r_\perp}{r_0}\right)^2\right] + Q(\infty),\ee
$A$ and $r_0$ being the fitting parameters. It is reasonable to treat $r_0$ as the radius of the confining string.
Function~(\ref{fit}) has been used in studies of the nonabelian
\cite{Bali:1994de} and abelian~\cite{Bali:1998de} flux 
tubes action density and predicted theoretically in~\cite{Luscher:1980iy}.
We obtained fits with reasonable $\chi^2$ (see Fig.~\ref{11x5}).

\begin{figure}
\psfrag{rrrrrr}{$r_0$, fm}
\psfrag{LLLLLL}{$\ln(R/1\,{\rm fm})$}

\epsfig{file=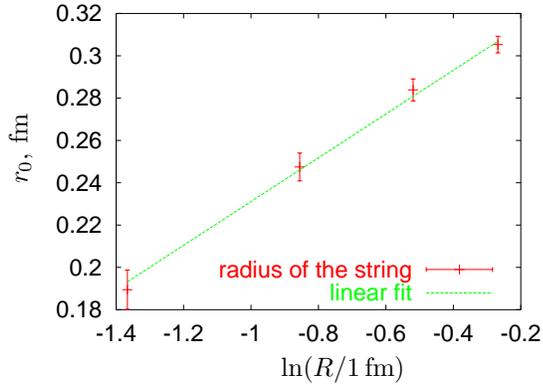,width=7.51cm}
\caption{$r_0$ as a function of $\ln R$, the result of the linear fit is
$r_0=0.104 \ln (R/1\,{\rm fm})+0.335\,\,({\rm fm}).$}
\label{r0}
\end{figure}
Fig.~\ref{r0} shows the parameter $r_0$ as a function of the distance between the quark and antiquark.
The radius of the string appears to be proportional
to the logarithm of the distance between the quark and antiquark
as predicted in~\cite{Luscher:1980iy} for the width of the nonabelian flux tube.
It is worth mentioning that the logarithmic dependence has been found out
in studies of the nonabelian flux tube~\cite{Bali:1994de},  while for the abelian 
flux tube the width was found to be stable~\cite{Bali:1998de}
(on the other hand, the accurate recent studies show the dependence of the width of the abelian flux tube on the distance between the test quark-antiquark pair~\cite{Koma}).
 
\section{Discussion}
For the monopole confinement mechanism there exists the effective classical model,
 which is the dual abelian Higgs model. The classical equations of motion for 
this model describe unexpectedly well the profile of the confining string 
in the maximal
abelian projection~\cite{AD}. For P-vortices we have no classical effective model, and our
fitting function (\ref{fit}) has a phenomenological origin.
On the other hand, 
the qualitative similarity of the dependence of $Q(r_\perp,r_\parallel)$ on $r_\perp$ to the dependence
of the action density of the nonabelian or abelian flux tube is not accidental.
$Q(r_\perp,r_\parallel)$ can be interpreted as $Z(2)$ projection of the nonabelian magnetic/electric energy density. This aspect will be discussed elsewhere 
\cite{bkps}.

We can also explain the flux tube profile shown in Fig.~\ref{3d1} from another point of view.
It is well known that monopoles are correlated with P-vortices~\cite{AB}. On the other hand,
it is also known that the condensate of monopoles is broken inside the 
confining string. Thus one can
expect that the condensate of P-vortices should be also  broken or at least 
substantially reduced inside the confining string.
We really see this effect in Figs.~\ref{3d1}--\ref{11x5}.

\section*{ACKNOWLEDGEMENTS}
The authors are grateful to M.~N.~Chernodub, F.~V.~Gubarev and V.~I.~Zakharov for useful discussions.
M.~I.~P. is partially supported by grants RFBR 02-02-17308, RFBR 01-02-117456, RFBR 00-15-96-786, INTAS-00-00111, and CRDF award RPI-2364-MO-02. A.~V.~K. is partially supported by grants
RFBR 02-02-17308 and CRDF MO-011-0. B.~V.~G. is supported by JSPS Fellowship
grant.

\end{document}